\documentclass[]{article}
\usepackage{amsmath}
\usepackage{amsfonts}
\usepackage{amssymb}
\usepackage{graphicx}
\usepackage{amsfonts}
\usepackage[english]{babel}
\usepackage[utf8]{inputenc}
\usepackage{times}
\usepackage[T1]{fontenc}

\begin{document}

\title{Analog of the Peter-Weyl  Expansion for Lorentz Group}

\author{Leonid Perlov\\
Department of Physics, University of Massachusetts,  Boston\\
leonid.perlov@umb.edu\\[2ex]
Dedicated to my teacher, Alexander Zibrov.
}

\maketitle

\begin{abstract}
The expansion of a square integrable function on $SL(2,C)$ into the sum of the principal series matrix coefficients with the specially selected representation parameters  was recently used in the Loop Quantum Gravity $\cite{RovelliBook2}$, $\cite{Rovelli2010}$. In this paper we prove that the sum $\sum\limits_{j=1}^{\infty}\sum\limits_{|m| \le j}\sum\limits_{|n| \le j} \frac{D^{(j, \tau j)}_{jm, jn}(g)}{j^k}$, where $ j, m, n \in Z, \tau \in C$ is convergent to a square integrable function on $SL(2,C)$. We also prove that for each fixed m:  $\sum\limits_{j=1}^{\infty}\frac{D^{(j, \tau j)}_{jm, jm}(g)}{j^k}$ is convergent and that the limit is a square integrable function on $SL(2,C)$.  We then prove convergence of the sums $\sum\limits_{j=|p|}^{\infty}\sum\limits_{|m| \le j}\sum\limits_{|n| \le j} d^{\frac{j}{2}}_{pm} D^{(j, \tau j)}_{jm, jn}(g)$, where $d^{\frac{j}{2}}_{|p|m} = (j+1)^{\frac{1}{2}}\int\limits_{SU(2)}\phi(u)\overline{ D^{\frac{j}{2}}_{|p|m}(u)} \; du \;  $  is $\phi(u)$'s Fourier transform and $ p, j, m, n \in Z, \tau \in C, u \in SU(2), g \in SL(2,C)$, thus establishing the map between the square integrable functions on $SU(2)$ and the space of the functions on $SL(2,C)$. Such maps were first used in $\cite{RovelliBook2}$.
\end{abstract}

\section{Introduction}

In this paper we prove a few facts about the sum of the Lorentz group matrix coefficients. The main result is the Theorem 3 establishing existence of the map from the square integrable functions on $SU(2)$ to the functions (not square integrable) on $SL(2,C)$: $\sum\limits_{j=|p|}^{\infty}\sum\limits_{|m| \le j}\sum\limits_{|n| \le j} d^{\frac{j}{2}}_{pm} D^{(j, \tau j)}_{jm, jn}(g)$\\
, where $d^{\frac{j}{2}}_{|p|m} = (j+1)^{\frac{1}{2}}\int\limits_{SU(2)}\phi(u)\overline{ D^{\frac{j}{2}}_{|p|m}(u)} \; du \;  $  is $\phi(u)$'s Fourier transform and $ p, j, m, n \in Z, \tau \in C, u \in SU(2), g \in SL(2,C)$. \\[2ex]
This map is called a Y-Map and plays a core role in the Loop Quantum Gravity. The map contains the summation not over all principal series parameters $(k, \rho)$, but rather over the parameters of the form:$(k = j, \rho = j\tau)$.\\
Such specific selection of the principal series coefficient parameters is not accidental. In fact $k = j, \rho = j \tau$ is the simplicity constraints solution. The simplicity constraints, introduced by John Barrett and Louse Crane in $\cite{BarrettCrain}$ allow to consider the Quantum Gravity as a 4-dimensional topological model called BF-model plus some constraints on the form of the bivectors used in BF model. Those constraints are called the simplicity constraints. The simplicity constraints is what makes the 4-dim topological model to become Einstein's Quantum Gravity. Thus the principal series matrix coefficients of the form $ D^{(j, j \tau)}_{jm, jm}$ have a special physical meaning. For the details please see $\cite{BarrettCrain}$, $\cite{FlippedVertex}$, $\cite{VertexAmplitude}$ and $\cite{EPRL}$.\\
In this paper the sum convergence proof is the main and the most challenging task. To prove convergence we use the following: a) the principal series matrix coefficients expression via the hypergeometric functions $\cite{DucHieu}$, b) formula (4.11), the Watson's asymptotic of the hypergeometric functions $_{2}F_{1}(a,b,c, z)$, when all three parameters tend to infinity $\cite{Watson}$, and c) the D'Alemebert-Cauchy convergence ratio test.\\
The paper is organized as follows. In the next section \ref{sec:PeterWeylExpansion1} we prove convergence of the sums $\sum\limits_{j=0}^{\infty}\sum\limits_{|m| \le j}\sum\limits_{|n| \le j} \frac{D^{(j, j \tau)}_{jm, jn}}{j^k}, \; k \ge 2$ and $\sum\limits_{j=m}^{\infty}\frac{D^{(j, j \tau)}_{jm, jm}}{j^k}, \; k \ge 2$.
In section \ref{sec:SquareIntegrability} we prove that the limits of the first  and second sums are square integrable functions. In the section \ref{sec:YMap} we establish the map from the space of the square integrable functions on $SU(2)$ to the space of functions on $SL(2,C)$ as the limit of Y-Map sums. In order to define such maps we prove the Y-Map sums convergence. The discussion section \ref{sec:Discussion} concludes the paper.  In Appendix we provide the graph from the numerical calculations by using MPMath Python program to demonstrate convergence proved in Lemma 1 visually.

\section{Convergence of the Sum of the Principal Series Matrix Coefficients}
\label{sec:PeterWeylExpansion1}

In this section we are going to prove two Lemmas stating that the following two sums are convergent:
$\sum\limits_{j=1}^{\infty}\sum\limits_{|m| \le j} \sum\limits_{|n| \le j}\frac{D^{(j, \tau j)}_{jm,  jn}(g)}{j^k}$ and $\sum\limits_{j=m}^{\infty}\frac{D^{(j, \tau j)}_{jm,  jm}(g)}{j^k}$ for $k \ge 2$  and fixed m, where $g \in SL(2,C), \; \tau \in C, j, n, m \in Z$. We will use these Lemmas in the following section in order to prove that both sums converge to the square integrable functions on $SL(2,C)$. \\[3ex]

\textbf{Lemma 1:} $\quad$ For $k \ge 2$ the sum $\sum\limits_{j=1}^{\infty}  \sum\limits_{|m| \le j} \frac{ D^{(j, \tau j)}_{jm,  jm}(g)}{j^k}$ is absolute convergent and therefore convergent for all $g \in SL(2,C), \; \tau \in C, j, m \in Z$. \\[2ex]
\textbf{Proof:}\\[2ex]
 Let us use the explicit expression for the matrix coefficients in ($\ref{DAlembert2}$). The first explicit expression of the principal series matrix coefficients $D^{(k', \rho)}_{jn, j'm}, k' \in Z, \rho \in C $ was obtained by Duc and Hieu in 1967  $\cite{DucHieu}$, formula (4.11):
\begin{multline}
\label{SL2CMatrix}
D^{(k', \rho)}_{jm, j'n}(g) =\frac{\delta_{mn}}{(j+j'+1)!}\\
{\left( (2j+1)(2j'+1)(j+m)!(j'+m)!(j-m)!(j'-m)!(j+k')!(j'+k')!(j-k')!(j'-k')! \right) }^{1/2}\\
\times \sum\limits_{d, d'} {(-1)}^{d + d'}\frac{(d+d'+m+k')!(j+j'-d-d'-m-k')!}{d!d'!(j-m-d)!(j'-m-d')!(k'+m+d)!(k' + m + d')!(j-k'-d)!(j'-k'-d')!}\\
\times {\epsilon}^{2(2d' + m + k' + 1 + \frac{i \rho}{2})}{}_2F_1( j' + 1 + \frac{i \rho}{2}, d+d'+m+k'+1; j+j'+2; 1-{\epsilon}^4)
\end{multline}
,where ${}_2F_1(\alpha, \beta; \gamma; z)$ - is a hypergeometric function, $d$ and $d'$ are integers that do not make each factor under the factorial to become a negative number and  $\epsilon$ is a real number obtained from the $g \in SL(2,C)$ decomposition:
\begin{equation}
\label{gdecomposition}
g=u_1 b u_2 
\end{equation}
,where $u_1$ and $u_2$ are unitary matrices, while the matrix $b$ =
$
\begin{pmatrix}
\label{matrix}
{\epsilon}^{-1}  & 0 \\
0 & \epsilon  \\
\end{pmatrix}
$, $\epsilon \in R$\\[2ex]
As one can see all $D^{(k', \rho)}_{jm, j'n}(g)$ are zero for $m \neq n$ due to the presence of the Kronecker delta in ($\ref{SL2CMatrix}$). Therefore we can omit all zero terms in the sums and leave only the terms with $m=n$. That's why we wrote our sum only over $j$ and $m$:  
\begin{equation}
\label{Psig2}
 \sum\limits_{j=1}^{\infty}\sum\limits_{|m| \le j} \frac{D^{(j, \tau j)}_{jm,  jm}(g)}{j^k}
\end{equation}
The matrix coefficients in our sum  $D^{(j, \tau j)}_{jm,  jm}(g)$ have much simpler form than the general form ($\ref{SL2CMatrix}$).
We rewrite the unitary matrix coefficients $D^{(k', \rho)}_{jm, j'n}(g)$ in ($\ref{SL2CMatrix}$) for:
$k' = j, \; \rho = \tau j,  \; j' = j, \; m=n$. Also since $d$ and $d'$ are so that factorial expressions are non-negative, one can see from  ($\ref{SL2CMatrix}$) that if $k' = j$, which is our case, then  $j-k' - d \ge 0$ implies $j-j - d \ge 0$, so $d \le 0$, but at the same time $d!$ implies $d \ge 0$ so it follows that $d=0$. The same is true for $d' = 0$ and the sums over $d$ and $d'$ in ($\ref{SL2CMatrix}$) disappear:
\begin{multline}
\label{SL2CMatrix1}
D^{(j, \tau j)}_{jm, jm}(g) =\frac{1}{(2j+1)!} (2j)!(2j+1)(j+m)!(j-m)!\times \frac{(j+m)!(j-m)!}{(j-m)!(j-m)!(j+m)!(j + m)!}\\
\times {\epsilon}^{2(m + j + 1 + \frac{i \tau j}{2})}{}_2F_1( j + 1 + \frac{i \tau j}{2}, m+j+1; 2j+2; 1-{\epsilon}^4)
\end{multline}
All coefficients cancel as one can see and we obtain:
\begin{equation}
\label{SL2CMatrix2}
D^{(j, \tau j)}_{jm, jm}(g) = {\epsilon}^{2(m + j + 1 + \frac{i \tau j}{2})} {}_2F_1( j + 1 + \frac{i \tau j}{2}, m+j+1; 2j+2; 1-{\epsilon}^4)
\end{equation}
The sum ($\ref{Psig2}$) becomes:
\begin{equation}
\label{NewSum}
\sum\limits_{j=1}^{\infty}\sum\limits_{|m| \le j}\frac{1}{j^k}  D^{(j, \tau j)}_{jm, jm}(g) = \sum\limits_{j=1}^{\infty}\sum\limits_{|m| \le j} \frac{1}{j^k} {\epsilon}^{2(m + j + 1 + \frac{i \tau j}{2})} {}_2F_1( j + 1 + \frac{i \tau j}{2}, m+j+1; 2j+2; 1-{\epsilon}^4) 
\end{equation}
We now consider the following two sums: first for $ 0 \le m \le j$  and the second for  $  -j \le m < 0 $ and by bounding them from above we will prove their convergence. The convergence of the original sum will then follow. 
\begin{multline}
\label{Bounding}
\sum\limits_{j=1}^{\infty}\frac{1}{j^k}\left | \sum\limits_{|m| \le j} D^{(j, \tau j)}_{jm, jm}(g) \right | \le \\
\sum\limits_{j=1}^{\infty} \frac{1}{j^k} \left | \sum\limits_{m = 0}^{m=j}  {\epsilon}^{2(m + j + 1 + \frac{i \tau j}{2})} {}_2F_1( j + 1 + \frac{i \tau j}{2}, m+j+1; 2j+2; 1-{\epsilon}^4)  \right | + \\
\sum\limits_{j=1}^{\infty} \frac{1}{j^k} \left | \sum\limits_{m = -j}^{m < 0}  {\epsilon}^{2(m + j + 1 + \frac{i \tau j}{2})} {}_2F_1( j + 1 + \frac{i \tau j}{2}, m+j+1; 2j+2; 1-{\epsilon}^4)  \right | \le \\
\sum\limits_{j=1}^{\infty}\frac{1}{j^k}\sum\limits_{m = 0}^{m=j} \left | {\epsilon}^{2(m + j + 1 + \frac{i \tau j}{2})} \right | \left |{}_2F_1( j + 1 + \frac{i \tau j}{2}, m+j+1; 2j+2; 1-{\epsilon}^4)  \right | + \\
\sum\limits_{j=1}^{\infty}\frac{1}{j^k}\sum\limits_{m = -j}^{m < 0} \left | {\epsilon}^{2(m + j + 1 + \frac{i \tau j}{2})} \right | \left |{}_2F_1( j + 1 + \frac{i \tau j}{2}, m+j+1; 2j+2; 1-{\epsilon}^4)  \right | \le \\
\sum\limits_{j=1}^{\infty} \frac{1}{j^k}  \left | (j+1) {\epsilon}^{2(j + j + 1 + \frac{i \tau j}{2})} \right | \left |{}_2F_1( j + 1 + \frac{i \tau j}{2}, j+j+1; 2j+2; 1-{\epsilon}^4)  \right | + \\
\sum\limits_{j=1}^{\infty} \frac{1}{j^k}  \left | j {\epsilon}^{2(0 + j + 1 + \frac{i \tau j}{2})} \right | \left |{}_2F_1( j + 1 + \frac{i \tau j}{2}, 0+j+1; 2j+2; 1-{\epsilon}^4)  \right |
\end{multline}
We pass to the last inequality above by putting $m = j$ in the first sum and $m = 0$ in the second and remembering  the hypergeometric function is monotonic with respect to its second argument:
\begin{equation}
\label{Hypergeometric}
{}_2F_1(a, b; c; z) = \sum\limits_{n=0}^{\infty} \frac{{(a)}_n {(b)}_n z^n}{{(c)}_n n!}\\
\end{equation}
,where 
\begin{equation}
{(q)}_n = 1, \mbox{when} \; n =0, {(q)}_n = q(q+1) \mbox{...} (q + n -1),   n > 0
\end{equation}
The hypergeometric function is originally defined for $|z| <1 $, but is analytically continued to all values of $z$ as was shown in $\cite{Watson}$.\\[2ex]
In our case of   ${}_2F_1( j + 1 + \frac{i \tau j}{2}, m+j+1; 2j+2; 1-{\epsilon}^4)$,  \;
the parameter $ b = m + j + 1 $ is always positive and the absolute value of the function is increasing when m is increasing. That is why in the last inequality of ($\ref{Bounding}$) we put $m = j$ to bound the sum from above when $ m \ge 0$ and by $m = 0$ in the second sum, when $m < 0$.\\[2ex]
At this point we are going to use the D'Alembert ratio convergence test and the asymptotic of the hypergeometric function to prove that the two bounding from above sums are convergent and that will prove that the original sum is convergent.
We will need to consider three cases: $|\epsilon| < 1$, $|\epsilon| > 1$, $\epsilon = 1$ \\[2ex]
The hypergeometric function ${}_2F_1(\alpha, \beta; \gamma; y)$ asymptotic, when all three parameters go to infinity, was investigated and derived by G.N Watson (1918) and can be found in Bateman's book $\cite{Bateman}$ volume 1 page 77:
\begin{multline}
\label{Asymptotic}
{\left(\frac{z}{2} -\frac{1}{2}\right)}^{-a -\lambda}{}_2F_1(a +\lambda, a - c + 1 + \lambda; a-b+1+2\lambda; 2{(1-z)}^{-1}) =\\ \frac{2^{a+b}\Gamma(a-b+1+2\lambda)\Gamma(1/2){\lambda}^{-1/2}}{\Gamma(a-c+1+\lambda)\Gamma(c-b+\lambda)} e^{-(a+\lambda)\xi}\times {(1-e^{-\xi})}^{-c+1/2} \times {(1+e^{-\xi})}^{c-a-b-1/2}[1+ O({\lambda}^{-1})]
\end{multline}
,where $\xi$ is defined as following: $e^{\pm \xi} = z \pm \sqrt{z^2-1}$. The minus sign corresponds to $Im(z) \le 0$, the plus sign to $Im(z) > 0$. This asymptotic also works in the limit case of z being real, which is our case of $1-{\epsilon}^4$ (for details see Watson's original 1918 paper $\cite{Watson}$) \\[2ex]
By comparing ($\ref{SL2CMatrix2}$) and $(\ref{Asymptotic})$  we see that the  hypergeometric function arguments $\lambda, a, b, c$  in our case take the following values:
\begin{equation}
\label{Parameters}
\lambda = j, \; a = 1 + \frac{i \tau j}{2}, \; b = \frac{i \tau j}{2}, \; c = 1+\frac{i \tau j}{2} - m, \; z = \frac{{\epsilon}^4 + 1}{{\epsilon}^4 -1}, \;  e^{\mp \xi} = \frac{{\epsilon}^2 \mp 1}{{\epsilon}^2 \pm 1}
\end{equation}
Indeed by substituting them into l.h.s of the ($\ref{Asymptotic}$) we get ${}_2F_1$ exactly as in $(\ref{SL2CMatrix2})$:\\[2ex]
${}_2F_1( j + 1 + \frac{i \tau j}{2}, m+j+1; 2j+2; 1-{\epsilon}^4)$\\[2ex]
Let us rewrite ($\ref{Asymptotic}$) then in terms of  $(j, m, \tau)$ and we obtain:
\begin{multline}
\label{Asymptotic1}
{}_2F_1( j + 1 + \frac{i \tau j}{2}, m+j+1; 2j+2; 1-{\epsilon}^4) = \\
\frac{1}{{({\epsilon}^4-1)}^{1+j+\frac{i \tau j}{2}}}
\frac{2^{(1+ i \tau j)}\Gamma(2 + 2j)\Gamma(\frac{1}{2})j^{-1/2}}{\Gamma(m+1+j) \Gamma(1-m+j)}\times \\
e^{-(1+\frac{i \tau j}{2} + j)\xi} \times  (1-e^{-\xi})^{(-\frac{1}{2}-\frac{i \tau j}{2}+m)} \times  (1+e^{-\xi})^{(-m -\frac{i \tau j}{2}-\frac{1}{2})}\left[ 1 + O(\frac{1}{j})\right]
\end{multline}
or by expressing $e^{-\xi}$ in terms of $\epsilon$ by using ($\ref{Parameters}$) we obtain the following expression:
\begin{multline}
\label{Asymptotic2}
{}_2F_1( j + 1 + \frac{i \tau j}{2}, m+j+1; 2j+2; 1-{\epsilon}^4) = \\
\frac{1}{{({\epsilon}^4-1)}^{1+j+\frac{i \tau j}{2}}}
\frac{2^{(1+ i \tau j)}\Gamma(2 + 2j)\Gamma(\frac{1}{2})j^{-1/2}}{\Gamma(m+1+j) \Gamma(1-m+j)}\times \\
\left ( {\frac{{\epsilon}^2 -1}{{\epsilon}^2 +1}}\right )^{(1+\frac{i \tau j}{2} + j)} \times  \left ( {\frac{2}{{\epsilon}^2 +1}} \right ) ^{(-\frac{1}{2}-\frac{i \tau j}{2}+m)}\times  \left ( {\frac{2{\epsilon}^2}{{\epsilon}^2 + 1}} \right ) ^{(-m -\frac{i \tau j}{2}-\frac{1}{2})}\left[ 1 + O(\frac{1}{j})\right]
\end{multline}
We rewrite this expression by denoting the right hand side before $\left[ 1 + O(\frac{1}{j})\right]$ as ${}_2A_1(j, m, \tau, \epsilon)$.
\begin{equation}
\label{Asymptotic22}
{}_2F_1( j + 1 + \frac{i \tau j}{2}, m+j+1; 2j+2; 1-{\epsilon}^4) ={}_2A_1(j, m, \tau, \epsilon) \left[ 1 + O(\frac{1}{j})\right]
\end{equation}
We are going to use this expression in the D'Alembert ratio test to prove the convergence of the bounding sums in ($\ref{Bounding}$). The first sum corresponds to $m = j$

\begin{multline}
\label{FirstSum}
\sum\limits_{j=1}^{\infty}\frac{1}{j^k} \left | (j+1) {\epsilon}^{2(j + j + 1 + \frac{i \tau j}{2})} \right | \left |{}_2F_1( j + 1 + \frac{i \tau j}{2}, j+j+1; 2j+2; 1-{\epsilon}^4)  \right | 
\end{multline}

while the second to $m=0$:

\begin{multline}
\label{SecondSum}
\sum\limits_{j=1}^{\infty} \frac{1}{j^k}  \left | j {\epsilon}^{2(0 + j + 1 + \frac{i \tau j}{2})} \right | \left |{}_2F_1( j + 1 + \frac{i \tau j}{2}, 0+j+1; 2j+2; 1-{\epsilon}^4)  \right |
\end{multline}
Before we begin, we can see right away that $ O(\frac{1}{j})$ in ($\ref{Asymptotic22}$) creates a problem for applying a ratio test. However it can be easily fixed. We substitute ($\ref{Asymptotic22}$)  into ($\ref{FirstSum}$)  and obtain:
\begin{multline}
\label{FirstSum1}
\sum\limits_{j=1}^{\infty}\frac{1}{j^k} \left | (j+1) {\epsilon}^{2(j + j + 1 + \frac{i \tau j}{2})} \right | \left|{}_2A_1(j, m, \tau, \epsilon)  \left[ 1 + O(\frac{1}{j})\right] \right| \le \\
\sum\limits_{j=1}^{\infty}\left | (j+1) {\epsilon}^{2(j + j + 1 + \frac{i \tau j}{2})} \right | \left|{}_2A_1(j, m, \tau, \epsilon) \right|\frac{1}{j^k}  + \\
\sum\limits_{j=1}^{\infty}\left | (j+1) {\epsilon}^{2(j + j + 1 + \frac{i \tau j}{2})} \right | \left|{}_2A_1(j, m, \tau, \epsilon) \right | \left[ O(\frac{1}{j^{k+1}})\right]  
\end{multline}
If we prove that $\sum\limits_{j=1}^{\infty}\left | (j+1) {\epsilon}^{2(j + j + 1 + \frac{i \tau j}{2})} \right | \left|{}_2A_1(j, m, \tau, \epsilon) \right|$ is convergent, then the first sum on the right hand side will be convergent for $k \ge 1$ by Abel's theorem, stating that if $\sum a_n$ is convergent sequence, and $b_n$ is monotonic bounded , then $\sum a_nb_n$ is convergent, while the second sum on the right will be convergent as $ O(\frac{1}{j^{k+1}})$ is convergent for $k \ge 1$, so it will be bounded by the product of two converging series: .
\begin{multline}
\sum\limits_{j=1}^{\infty}\left | (j+1) {\epsilon}^{2(j + j + 1 + \frac{i \tau j}{2})} \right | \left|{}_2A_1(j, m, \tau, \epsilon) \right | \left[ O(\frac{1}{j^{k+1}})\right]  \le \\
 \sum\limits_{j=1}^{\infty}\left | (j+1) {\epsilon}^{2(j + j + 1 + \frac{i \tau j}{2})} \right | \left|{}_2A_1(j, m, \tau, \epsilon) \right |  \times  \sum\limits_{j=1}^{\infty}\left | \left[ O(\frac{1}{j^{k+1}})\right] \right|
\end{multline} 
The same logic applies to the second bounding sum ($\ref{SecondSum}$). \\
Therefore, in order to prove the bounding sums  ($\ref{FirstSum}$), ($\ref{SecondSum}$) convergence it is enough to prove the convergence of:
\begin{equation}
\label{FirstSum3}
\sum\limits_{j=1}^{\infty}\left | (j+1) {\epsilon}^{2(j + j + 1 + \frac{i \tau j}{2})} \right | \left|{}_2A_1(j, m, \tau, \epsilon) \right|
\end{equation}
and
\begin{equation}
\label{SecondSum3}
\sum\limits_{j=1}^{\infty}  \left | j {\epsilon}^{2(0 + j + 1 + \frac{i \tau j}{2})} \right | \left |{}_2A_1(j,m, \tau, \epsilon )  \right |
\end{equation}
, where we remind that by definition from ($\ref{Asymptotic2}$)
\begin{multline}
\label{ADefinition}
{}_2A_1(j,m, \tau, \epsilon ) = {}_2A_1( j + 1 + \frac{i \tau j}{2}, m+j+1; 2j+2; 1-{\epsilon}^4) = \\
 \frac{1}{{({\epsilon}^4-1)}^{1+j+\frac{i \tau j}{2}}}
\frac{2^{(1+ i \tau j)}\Gamma(2 + 2j)\Gamma(\frac{1}{2})j^{-1/2}}{\Gamma(m+1+j) \Gamma(1-m+j)}\times \\
\left ( {\frac{{\epsilon}^2 -1}{{\epsilon}^2 +1}}\right )^{(1+\frac{i \tau j}{2} + j)} \times  \left ( {\frac{2}{{\epsilon}^2 +1}} \right ) ^{(-\frac{1}{2}-\frac{i \tau j}{2}+m)}\times  \left ( {\frac{2{\epsilon}^2}{{\epsilon}^2 + 1}} \right ) ^{(-m -\frac{i \tau j}{2}-\frac{1}{2})}
\end{multline}
We used two notations for the same function ${}_2A_1$ above. The second one is for showing the arguments explicitly, which is more convenient, when we begin using the ratio tests below.\\
By proving the sums convergence we would need to consider two cases of $|\epsilon| > 1$ and $|\epsilon| < 1$ for each sum separately, i.e. four cases all together. The simple fifth case $\epsilon = 1$ is considered at the end.\\[2ex]
\textbf{Case 1:}  First sum, $m = j$, $\tau \in C, \tau = \eta + i \omega$, $|\epsilon| > 1$\\[2ex]
\begin{equation}
\sum\limits_{j=1}^{\infty} \left | (j+1) {\epsilon}^{2(2j + 1 + \frac{i \tau j}{2})} \right | \left |{}_2A_1( j + 1 + \frac{i \tau j}{2}, 2j+1; 2j+2; 1-{\epsilon}^4)  \right |
\end{equation}
The D'Alembert ratio test is as follows:
\begin{multline}
\lim\limits_{j \rightarrow \infty} \left | \frac{(j+2) {\epsilon}^{2(2(j+1) + 1 + \frac{i \tau (j+1)}{2})}}{(j+1) {\epsilon}^{2(2j+ 1 + \frac{i \tau j}{2})}} \right | \left |\frac{{}_2A_1( j + 2 + \frac{i \tau (j+1)}{2}, 2(j+1)+1; 2(j+1)+2; 1-{\epsilon}^4)}{{}_2A_1( j + 1 + \frac{i \tau j}{2}, 2j+1; 2j+2; 1-{\epsilon}^4)}  \right | = \\
{\epsilon}^4 {\epsilon}^{-\omega}\times \lim\limits_{j \rightarrow \infty} \left |\frac{{}_2A_1( j + 2 + \frac{i \tau (j+1)}{2}, 2(j+1)+1; 2(j+1)+2; 1-{\epsilon}^4)}{{}_2A_1( j + 1 + \frac{i \tau j}{2}, 2j+1; 2j+2; 1-{\epsilon}^4)} \right |
\end{multline}
by using  ($\ref{ADefinition}$) for  $m = j$ we obtain:
\begin{multline}
 \lim\limits_{j \rightarrow \infty} {\epsilon}^4 {\epsilon}^{-\omega}\times \left |\frac{{}_2A_1( j + 2 + \frac{i \tau (j+1)}{2}, 2(j+1)+1; 2(j+1)+2; 1-{\epsilon}^4)}{{}_2A_1( j + 1 + \frac{i \tau j}{2}, 2j+1; 2j+2; 1-{\epsilon}^4)} \right | = \\
 \lim\limits_{j \rightarrow \infty} {\epsilon}^4 {\epsilon}^{-\omega}\times \left |  \frac{{({\epsilon}^4-1)}^{1+j+\frac{i \tau j}{2}}}{{({\epsilon}^4-1)}^{1+j+1+\frac{i \tau (j+1)}{2}}}\frac{2^{(1+i\tau (j+1))}}{2^{(1+i\tau j)}}\frac{\Gamma(2 + 2(j+1))\Gamma(\frac{1}{2}){(j+1)}^{-1/2}\Gamma(2j+1)\Gamma(1)}{\Gamma(2+2j)\Gamma(\frac{1}{2})j^{-1/2}\Gamma(2(j+1)+1) \Gamma(1)} \right | \times  \\
\left | \left ( {\frac{{\epsilon}^2 -1}{{\epsilon}^2 +1}}\right )^{(1+\frac{i \tau (j+1)}{2} + (j+1))-(1+\frac{i \tau j}{2} + j)} \right |  \times \left | \left ( {\frac{2}{{\epsilon}^2 +1}} \right ) ^{(-\frac{1}{2}-\frac{i \tau (j+1)}{2}+(j+1))-(-\frac{1}{2}-\frac{i \tau j}{2}+j)} \right |  \\
\times  \left | \left ( {\frac{2{\epsilon}^2}{{\epsilon}^2 + 1}} \right ) ^{(-j -1 -\frac{i \tau (j+1)}{2}-\frac{1}{2})-(-j -\frac{i \tau j}{2}-\frac{1}{2})} \right | = \\
  \lim\limits_{j \rightarrow \infty}  \left | \frac{{\epsilon}^4 {\epsilon}^{-\omega} 2^{-\omega}}{({\epsilon}^4-1)^{(1-\frac{\omega}{2})}}   \frac{(2j+3)(2j+2)\Gamma(2j+2)\Gamma(2j+1)}{(2j+2)(2j+1)\Gamma(2j+2)\Gamma(2j+1)}    \frac{{({\epsilon}^2 -1)}^{(1-\frac{\omega}{2})}}{{({\epsilon}^2 +1)}^{(1-\frac{\omega}{2})}}\frac{2^{(1+\frac{\omega}{2})}}{{({\epsilon}^2+1)}^{(1+\frac{\omega}{2})}} \frac{{({\epsilon}^2 + 1)}^{(1-\frac{\omega}{2})}}{{(2{\epsilon}^2)}^{(1-\frac{\omega}{2})}} \right | = \\
\frac{{\epsilon}^2}{{({\epsilon}^2 + 1)}^2} < 1,  \;  \forall |\epsilon| > 1 
\end{multline}
 We used the fact that the absolute value of the positive real number in the pure imaginary power is 1 and the property of the $\Gamma$ function: $\Gamma(z+1) = z \Gamma(z)$. By this property all $\Gamma$ above cancel. We also remind that $\omega$ in the formula above comes from $\tau = \eta + i \omega$. \\[2ex]

\textbf{Case 2:}  First sum $m = j$, $\tau \in C, \tau = \eta + i \omega$, $|\epsilon| < 1$\\[2ex]
\begin{equation}
\sum\limits_{j=1}^{\infty} \left | (j+1) {\epsilon}^{2(2j + 1 + \frac{i \tau j}{2})} \right | \left |{}_2A_1( j + 1 + \frac{i \tau j}{2}, 2j+1; 2j+2; 1-{\epsilon}^4)  \right |
\end{equation}
The D'Alembert ratio test provides the expression very similar to the Case 1 with one difference. In this case of $|\epsilon| < 1$
we write the following expressions in the form:
\begin{equation}
 {\epsilon}^4 - 1 = (1 - {\epsilon}^4) e^{\pm i \pi}
\end{equation} 
\begin{equation}
{\epsilon}^2 - 1 = (1 - {\epsilon}^2) e^{\pm i \pi}
\end{equation}
\begin{multline}
 \lim\limits_{j \rightarrow \infty} {\epsilon}^4 {\epsilon}^{-\omega} \times \left |\frac{{}_2A_1( j + 2 + \frac{i \tau (j+1)}{2}, 2(j+1)+1; 2(j+1)+2; 1-{\epsilon}^4)}{{}_2A_1( j + 1 + \frac{i \tau j}{2}, 2j+1; 2j+2; 1-{\epsilon}^4)} \right | = \\
 \lim\limits_{j \rightarrow \infty} {\epsilon}^4 {\epsilon}^{-\omega}\times \left |  \frac{{(({1-\epsilon}^4) e^{\pm i \pi})}^{1+j+\frac{i \tau j}{2}}}{{(({1-\epsilon}^4) e^{\pm i \pi}})^{1+j+1+\frac{i \tau (j+1)}{2}}}\frac{2^{(1+i\tau (j+1))}}{2^{(1+i\tau j)}} \frac{\Gamma(2 + 2(j+1))\Gamma(\frac{1}{2}){(j+1)}^{-1/2}\Gamma(2j+1)\Gamma(1)}{\Gamma(2+2j)\Gamma(\frac{1}{2})j^{-1/2}\Gamma(2(j+1)+1) \Gamma(1)} \right | \times  \\
\left | \left ( {\frac{(1-{\epsilon}^2) e^{\pm i \pi}}{{\epsilon}^2 +1}}\right )^{(1+\frac{i \tau (j+1)}{2} + (j+1))-(1+\frac{i \tau j}{2} + j)} \right | \times  \left | \left ( {\frac{2}{{\epsilon}^2 +1}} \right ) ^{(-\frac{1}{2}-\frac{i \tau (j+1)}{2}+(j+1))-(-\frac{1}{2}-\frac{i \tau j}{2}+j)} \right | \times \\ 
\left | \left ( {\frac{2{\epsilon}^2}{{\epsilon}^2 + 1}} \right ) ^{(-j -1 -\frac{i \tau (j+1)}{2}-\frac{1}{2})-(-j -\frac{i \tau j}{2}-\frac{1}{2})} \right | = \\
	  \lim\limits_{j \rightarrow \infty}  \left | \frac{{\epsilon}^4 {\epsilon}^{-\omega}2^{-\omega}}{(1-{\epsilon}^4)^{(1-\frac{\omega}{2})} e^{\mp \frac{\pi \eta}{2}}}  \frac{(2j+3)(2j+2)\Gamma(2j+2)\Gamma(2j+1)}{(2j+2)(2j+1)\Gamma(2j+2)\Gamma(2j+1)}    \frac{(1-{\epsilon}^2)^{(1-\frac{\omega}{2})}e^{\mp \frac{\pi \eta}{2}}}{({\epsilon}^2 +1)^{(1-\frac{\omega}{2})}}\frac{2^{(1+\frac{\omega}{2})}}{({\epsilon}^2+1)^{(1+\frac{\omega}{2})}} \frac{({\epsilon}^2 + 1)^{(1-\frac{\omega}{2})}}{(2{\epsilon}^2)^{(1-\frac{\omega}{2})}} \right | = \\ 
\frac{{\epsilon}^2}{{({\epsilon}^2 + 1)}^2} < 1, \; \forall |\epsilon| < 1
\end{multline}
\textbf{Case 3:}  Second sum,  $m = 0$, $\tau \in C, \tau = \eta + i \omega $, $|\epsilon| > 1$\\[2ex]
\begin{equation}
\sum\limits_{j=1}^{\infty} \left | j {\epsilon}^{2(j + 1 + \frac{i \tau j}{2})} \right | \left |{}_2A_1( j + 1 + \frac{i \tau j}{2}, j+1; 2j+2; 1-{\epsilon}^4)  \right |
\end{equation}
D'Alembert ratio test is as follows:
\begin{multline}
\lim\limits_{j \rightarrow \infty} \left | \frac{(j+1) {\epsilon}^{2(j+1+ 1 + \frac{i \tau (j+1)}{2})}}{j {\epsilon}^{2(j + 1 + \frac{i \tau j}{2})}} \right | \left |\frac{{}_2A_1( j + 2+ \frac{i \tau (j+1)}{2}, j+2; 2(j+1)+2; 1-{\epsilon}^4)}{{}_2A_1( j + 1 + \frac{i \tau j}{2}, j+1; 2j+2; 1-{\epsilon}^4)}  \right | = \\
{\epsilon}^2 {\epsilon}^{-\omega}\times \lim\limits_{j \rightarrow \infty} \left |\frac{{}_2A_1( j + 2+ \frac{i \tau (j+1)}{2}, j+2; 2(j+1)+2; 1-{\epsilon}^4)}{{}_2A_1( j + 1 + \frac{i \tau j}{2}, j+1; 2j+2; 1-{\epsilon}^4)}  \right |
\end{multline}
We use ($\ref{ADefinition}$) for $m = 0$ 
\begin{multline}
 \lim\limits_{j \rightarrow \infty}  {\epsilon}^2 {\epsilon}^{-\omega}\left |\frac{{}_2A_1( j + 2+ \frac{i \tau (j+1)}{2}, j+2; 2(j+1)+2; 1-{\epsilon}^4)}{{}_2A_1( j + 1 + \frac{i \tau j}{2}, j+1; 2j+2; 1-{\epsilon}^4)}  \right | = \\
 \lim\limits_{j \rightarrow \infty} {\epsilon}^2 {\epsilon}^{-\omega} \times \left |  \frac{{({\epsilon}^4-1)}^{1+j+\frac{i \tau j}{2}}}{{({\epsilon}^4-1)}^{1+j+1+\frac{i \tau (j+1)}{2}}}\frac{2^{(1+i\tau (j+1))}}{2^{(1+i\tau j)}} \frac{\Gamma(2 + 2(j+1))\Gamma(\frac{1}{2}){(j+1)}^{-1/2}\Gamma(j+1)\Gamma(j+1)}{\Gamma(2+2j)\Gamma(\frac{1}{2})j^{-1/2}\Gamma(j+2) \Gamma(j+2)} \right | \times  \\
\left | \left ( {\frac{{\epsilon}^2 -1}{{\epsilon}^2 +1}}\right )^{(1+\frac{i \tau (j+1)}{2} + (j+1))-(1+\frac{i \tau j}{2} + j)} \right | \times \left | \left ( {\frac{2}{{\epsilon}^2 +1}} \right ) ^{(-\frac{1}{2}-\frac{i \tau (j+1)}{2})-(-\frac{1}{2}-\frac{i \tau j}{2})} \right | \\
 \times \left | \left ( {\frac{2{\epsilon}^2}{{\epsilon}^2 + 1}} \right ) ^{( -\frac{i \tau (j+1)}{2}-\frac{1}{2})-( -\frac{i \tau j}{2}-\frac{1}{2})} \right | = \\
  \lim\limits_{j \rightarrow \infty}  \left | \frac{{\epsilon}^2{\epsilon}^{-\omega}2^{-\omega}}{({\epsilon}^4-1)^{(1-\frac{\omega}{2})}}  \frac{(2j+3)(2j+2)\Gamma(2j+2)\Gamma(j+1)\Gamma(j+1)}{(j+1)(j+1)\Gamma(2j+2)\Gamma(j+1)\Gamma(j+1)}    \frac{({\epsilon}^2 -1)^{(1-\frac{\omega}{2})}}{({\epsilon}^2 +1)^{(1-\frac{\omega}{2})}} \frac{2^{\frac{\omega}{2}}}{({\epsilon}^2 + 1)^{\frac{\omega}{2}}} \frac{({2{\epsilon}^2)}^{\frac{\omega}{2}}}{({\epsilon}^2 + 1)^{\frac{\omega}{2}}} \right | = \\
\frac{4{\epsilon}^2}{{({\epsilon}^2 + 1)}^2} < 1, \;  \forall |\epsilon| > 1
\end{multline}

\textbf{Case 4:}  Second sum, $m = 0$, $\tau \in C, \tau = \eta + i \omega,  |\epsilon| < 1$\\[2ex]
\begin{equation}
\sum\limits_{j=1}^{\infty} \left | j {\epsilon}^{2(j + 1 + \frac{i \tau j}{2})} \right | \left | {}_2A_1( j + 1 + \frac{i \tau j}{2}, j+1; 2j+2; 1-{\epsilon}^4)\right |
\end{equation}
This case is similar to Case 3. For $|\epsilon| < 1$ we need to write again the following two expressions in the form:
\begin{equation}
 {\epsilon}^4 - 1 = (1 - {\epsilon}^4) e^{\pm i \pi}
\end{equation} 
\begin{equation}
{\epsilon}^2 - 1 = (1 - {\epsilon}^2) e^{\pm i \pi}
\end{equation}

The D'Alembert ratio test is as follows:
\begin{multline}
\lim\limits_{j \rightarrow \infty} \left | \frac{(j+1) {\epsilon}^{2(j+1+ 1 + \frac{i \tau (j+1)}{2})}}{j {\epsilon}^{2(j + 1 + \frac{i \tau j}{2})}} \right | \left |\frac{{}_2A_1( j + 2+ \frac{i \tau (j+1)}{2}, j+2; 2(j+1)+2; 1-{\epsilon}^4)}{{}_2A_1( j + 1 + \frac{i \tau j}{2}, j+1; 2j+2; 1-{\epsilon}^4)}  \right | = \\
{\epsilon}^2 {\epsilon}^{-\omega} \times \lim\limits_{j \rightarrow \infty} \left |\frac{{}_2A_1( j + 2+ \frac{i \tau (j+1)}{2}, j+2; 2(j+1)+2; 1-{\epsilon}^4)}{{}_2A_1( j + 1 + \frac{i \tau j}{2}, j+1; 2j+2; 1-{\epsilon}^4)}  \right |
\end{multline}
By using  ($\ref{ADefinition}$) for $m=0$ and $|\epsilon| < 1$
\begin{multline}
 \lim\limits_{j \rightarrow \infty}  {\epsilon}^2 {\epsilon}^{-\omega} \left |\frac{{}_2A_1( j + 2+ \frac{i \tau (j+1)}{2}, j+2; 2(j+1)+2; 1-{\epsilon}^4)}{{}_2A_1( j + 1 + \frac{i \tau j}{2}, j+1; 2j+2; 1-{\epsilon}^4)}  \right | = \\
 \lim\limits_{j \rightarrow \infty} {\epsilon}^2 {\epsilon}^{-\omega} \times \left |  \frac{({(1-{\epsilon}^4)e^{\pm i \pi})}^{1+j+\frac{i \tau j}{2}}}{{(({1-\epsilon}^4) e ^{\pm i \pi})}^{1+j+1+\frac{i \tau (j+1)}{2}}}\frac{2^{(1+i\tau (j+1))}}{2^{(1+i\tau j)}} \frac{\Gamma(2 + 2(j+1))\Gamma(\frac{1}{2}){(j+1)}^{-1/2}\Gamma(j+1)\Gamma(j+1)}{\Gamma(2+2j)\Gamma(\frac{1}{2})j^{-1/2}\Gamma(j+2) \Gamma(j+2)} \right | \times  \\
\left | \left ( {\frac{(1-{\epsilon}^2) e^{\pm i \pi} }{{\epsilon}^2 +1}}\right )^{(1+\frac{i \tau (j+1)}{2} + (j+1))-(1+\frac{i \tau j}{2} + j)} \right | \times  \left | \left ( {\frac{2}{{\epsilon}^2 +1}} \right ) ^{(-\frac{1}{2}-\frac{i \tau (j+1)}{2})-(-\frac{1}{2}-\frac{i \tau j}{2})} \right | \\
\times \left | \left ( {\frac{2{\epsilon}^2}{{\epsilon}^2 + 1}} \right ) ^{( -\frac{i \tau (j+1)}{2}-\frac{1}{2})-( -\frac{i \tau j}{2}-\frac{1}{2})} \right | = \\
  \lim\limits_{j \rightarrow \infty}  \left | \frac{{\epsilon}^2 {\epsilon}^{-\omega} 2^{-{\omega}}}{(1-{\epsilon}^4)^{(1-\frac{\omega}{2})}e^{\mp \frac{\pi \eta}{2}}}  \frac{(2j+3)(2j+2)\Gamma(2j+2)\Gamma(j+1)\Gamma(j+1)}{(j+1)(j+1)\Gamma(2j+2)\Gamma(j+1)\Gamma(j+1)}    \frac{(1-{\epsilon}^2 )^{(1-\frac{\omega}{2})} e^{\mp \frac{\pi \eta}{2}}}{({\epsilon}^2 +1)^{(1-\frac{\omega}{2})}}  \frac{2^{\frac{\omega}{2}}}{({\epsilon}^2 + 1)^{\frac{\omega}{2}}} \frac{({2{\epsilon}^2)}^{\frac{\omega}{2}}}{({\epsilon}^2 + 1)^{\frac{\omega}{2}}} \right | =\\ \frac{4{\epsilon}^2}{{({\epsilon}^2 + 1)}^2} < 1, \quad \forall |\epsilon| < 1 
\end{multline}
\textbf{Case 5:} $\epsilon = 1$\\
In the remaining case $\epsilon = 1$ we get $ {}_2F_1(a, b; c; 0) = 1 $ and $D^{(j, \tau j)}_{jm,  jm}(g) =1$ therefore 
$\sum\limits_{j=1}^{\infty}\sum\limits_{|m| \le j} \frac{1}{j^k}D^{(j, \tau j)}_{jm,  jm}(g) =\sum\limits_{j=1}^{\infty} \frac{2j+1}{j^k} $ is convergent for $k \ge 2$, as it is a Riemann zeta function.\\[2ex]
Thus, we have proved by ratio test that the limiting sums ($\ref{FirstSum3}$) and ($\ref{SecondSum3}$) are convergent, and hence by ($\ref{FirstSum1}$) trick, the limiting sums ($\ref{FirstSum}$) and ($\ref{SecondSum}$) are convergent, and therefore by ($\ref{Bounding}$) $\sum\limits_{j=1}^{\infty}  \sum\limits_{|m| \le j} \frac{ D^{(j, \tau j)}_{jm,  jm}(g)}{j^k}$ is absolute convergent and therefore convergent for all $g \in SL(2,C), \; \tau \in C, j, m \in Z$. \\[2ex]

$\square$\\[3ex]
\textbf{Lemma 2:} $\quad$ The sum $\sum\limits_{j=1}^{\infty} \frac{D^{(j, \tau j)}_{jm,  jm}(g)}{j^k}, k \ge 2 $ is absolute convergent  and therefore convergent for each $m \in Z$, $\forall g \in SL(2,C), \; \tau \in C, j \in Z$. \\[2ex]
\textbf{Proof:}\\[2ex]
The proof is very similar to the proof of the Lemma 1. We need to consider only three cases $|\epsilon| > 1, |\epsilon| < 1$ and $|\epsilon| = 1$, instead of five cases of the Lemma 1. This is due to the absence of the sums over m and n and therefore there is no need in the bounding sums.\\[2ex]
By using the explicit form of the matrix coefficients ($\ref{SL2CMatrix2}$) and the asymptotic ($\ref{Asymptotic2}$) we can write the asymptotic of the matrix coefficients in the form:
\begin{multline}
\label{Asymptotic3}
D^{(j, \tau j)}_{jm, jm}(g) = {\epsilon}^{2(m + j + 1 + \frac{i \tau j}{2})} {}_2F_1( j + 1 + \frac{i \tau j}{2}, m+j+1; 2j+2; 1-{\epsilon}^4) = \\
{\epsilon}^{2(m + j + 1 + \frac{i \tau j}{2})} \frac{1}{{({\epsilon}^4-1)}^{1+j+\frac{i \tau j}{2}}}
\frac{2^{(1+ i \tau j)}\Gamma(2 + 2j)\Gamma(\frac{1}{2})j^{-1/2}}{\Gamma(m+1+j) \Gamma(1-m+j)}\times \\
\left ( {\frac{{\epsilon}^2 -1}{{\epsilon}^2 +1}}\right )^{(1+\frac{i \tau j}{2} + j)} \times  \left ( {\frac{2}{{\epsilon}^2 +1}} \right ) ^{(-\frac{1}{2}-\frac{i \tau j}{2}+m)}\times  \left ( {\frac{2{\epsilon}^2}{{\epsilon}^2 + 1}} \right ) ^{(-m -\frac{i \tau j}{2}-\frac{1}{2})}\left[ 1 + O(\frac{1}{j})\right]
\end{multline}
As in Lemma1 we introduce the following notation:
\begin{equation}
\label{Asymptotic23}
{}_2F_1( j + 1 + \frac{i \tau j}{2}, m+j+1; 2j+2; 1-{\epsilon}^4) ={}_2A_1(j, m, \tau, \epsilon) \left[ 1 + O(\frac{1}{j})\right]
\end{equation}
Then by substituting ($\ref{Asymptotic23}$) into ($\ref{Asymptotic3}$):
\begin{multline}
\label{FirstSum11}
\sum\limits_{j=1}^{\infty} \frac{D^{(j, \tau j)}_{jm,  jm}(g)}{j^k}=\sum\limits_{j=1}^{\infty}\frac{1}{j^k} \left |  {\epsilon}^{2(m + j + 1 + \frac{i \tau j}{2})} \right | \left|{}_2A_1(j, m, \tau, \epsilon)  \left[ 1 + O(\frac{1}{j})\right] \right| \le \\
\sum\limits_{j=1}^{\infty}\left | {\epsilon}^{2(m + j + 1 + \frac{i \tau j}{2})} \right | \left|{}_2A_1(j, m, \tau, \epsilon) \right|\frac{1}{j^k}  + \\
\sum\limits_{j=1}^{\infty}\left |  {\epsilon}^{2(m + j + 1 + \frac{i \tau j}{2})} \right | \left|{}_2A_1(j, m, \tau, \epsilon) \right | \left[ O(\frac{1}{j^{k+1}})\right]  
\end{multline}
If we prove that $\sum\limits_{j=1}^{\infty}\left |  {\epsilon}^{2(m + j + 1 + \frac{i \tau j}{2})} \right | \left|{}_2A_1(j, m, \tau, \epsilon) \right|$ is convergent, then the first sum on the right hand side will be convergent for $k \ge 1$ by Abel's theorem, stating that if $\sum a_n$ is convergent sequence, and $b_n$ is monotonic bounded , then $\sum a_nb_n$ is convergent, while the second sum on the right will be convergent as $ O(\frac{1}{j^{k+1}})$ is convergent for $k \ge 1$, so it will be bounded by the product of two converging series: .
\begin{multline}
\label{FirstSum12}
\sum\limits_{j=1}^{\infty}\left | {\epsilon}^{2(m + j + 1 + \frac{i \tau j}{2})} \right | \left|{}_2A_1(j, m, \tau, \epsilon) \right | \left[ O(\frac{1}{j^{k+1}})\right]  \le \\
 \sum\limits_{j=1}^{\infty}\left |{\epsilon}^{2(m + j + 1 + \frac{i \tau j}{2})} \right | \left|{}_2A_1(j, m, \tau, \epsilon) \right |  \times  \sum\limits_{j=1}^{\infty}\left | \left[ O(\frac{1}{j^{k+1}})\right] \right|
\end{multline} 
We use D'Alembert ratio test to prove that 
 $\sum\limits_{j=1}^{\infty}\left |  {\epsilon}^{2(m + j + 1 + \frac{i \tau j}{2})} \right | \left|{}_2A_1(j, m, \tau, \epsilon) \right|$ is convergent:\\[2ex]
\textbf{Case 1:} $|\epsilon| > 1, \tau \in C, \tau = \eta + i \omega$
\begin{multline}
\label{DAlembert3}
\lim\limits_{j \rightarrow \infty} \left |  \frac{{\epsilon}^{2(m + j + 2 + \frac{i \tau (j+1)}{2})} {}_2A_1( j + 2 + \frac{i \tau (j+1)}{2}, m+j+2; 2j+4; 1-{\epsilon}^4) }{{\epsilon}^{2(m + j + 1 + \frac{i \tau j}{2})} {}_2A_1( j + 1 + \frac{i \tau j}{2}, m+j+1; 2j+2; 1-{\epsilon}^4)} \right | = \\
\lim\limits_{j \rightarrow \infty} {\epsilon}^2 {\epsilon}^{-\omega}\times \left |  \frac{{({\epsilon}^4-1)}^{1+j+\frac{i \tau j}{2}}}{{({\epsilon}^4-1)}^{1+j+1+\frac{i \tau (j+1)}{2}}}\frac{2^{(1+i\tau (j+1))}}{2^{(1+i\tau j)}}\frac{\Gamma(2 + 2(j+1))\Gamma(\frac{1}{2}){(j+1)}^{-1/2}\Gamma(m+1+j)\Gamma(1-m+j)}{\Gamma(2+2j)\Gamma(\frac{1}{2})j^{-1/2}\Gamma(m+2+j) \Gamma(2-m+j)} \right | \times  \\
\left | \left ( {\frac{{\epsilon}^2 -1}{{\epsilon}^2 +1}}\right )^{(1+\frac{i \tau (j+1)}{2} + j+1)-(1+\frac{i \tau j}{2} + j)} \right |  \times \left | \left ( {\frac{2}{{\epsilon}^2 +1}} \right ) ^{(-\frac{1}{2}-\frac{i \tau (j+1)}{2}+m)-(-\frac{1}{2}-\frac{i \tau j}{2}+m)} \right |  \\
\times  \left | \left ( {\frac{2{\epsilon}^2}{{\epsilon}^2 + 1}} \right ) ^{(-m -\frac{i \tau (j+1)}{2}-\frac{1}{2})-(-m -\frac{i \tau j}{2}-\frac{1}{2})} \right | = \\
  \lim\limits_{j \rightarrow \infty}  \left | \frac{{\epsilon}^2 {\epsilon}^{-\omega} 2^{-\omega}}{({\epsilon}^4-1)^{(1-\frac{\omega}{2})}}   \frac{(2j+3)(2j+2)\Gamma(2j+2)\Gamma(m+1+j)\Gamma(1-m+j)}{\Gamma(2j+2)(m+1+j)(1-m+j)\Gamma(m+1+j)\Gamma(1-m+j)} \right |  \times \\ \left |  \frac{{({\epsilon}^2 -1)}^{(1-\frac{\omega}{2})}}{{({\epsilon}^2 +1)}^{(1-\frac{\omega}{2})}}\frac{2^{\frac{\omega}{2}}}{{({\epsilon}^2+1)}^{\frac{\omega}{2}}} \frac{{(2{\epsilon}^2)}^{(\frac{\omega}{2})}}{{({\epsilon}^2 + 1)}^{(\frac{\omega}{2})}} \right | = 
\frac{4{\epsilon}^2}{{({\epsilon}^2 + 1)}^2} < 1,  \; \forall |\epsilon| > 1 
\end{multline}\\[2ex]
The Case 2 is very similar to the Case 1; the only difference is that, when $|\epsilon| < 1$ we need to write the following expressions in the form:
\begin{equation}
 {\epsilon}^4 - 1 = (1 - {\epsilon}^4) e^{\pm i \pi}
\end{equation} 
\begin{equation}
{\epsilon}^2 - 1 = (1 - {\epsilon}^2) e^{\pm i \pi}
\end{equation}
\textbf{Case 2:} $|\epsilon| < 1 \; , \tau \in C, \tau = \eta + i \omega$
\begin{multline}
\label{DAlembert4}
 \lim\limits_{j \rightarrow \infty}  
\left |  \frac{{\epsilon}^{2(m + j + 2 + \frac{i \tau (j+1)}{2})} {}_2A_1( j + 2 + \frac{i \tau (j+1)}{2}, m+j+2; 2j+4; 1-{\epsilon}^4) }{{\epsilon}^{2(m + j + 1 + \frac{i \tau j}{2})} {}_2A_1( j + 1 + \frac{i \tau j}{2}, m+j+1; 2j+2; 1-{\epsilon}^4)} \right | = \\
\lim\limits_{j \rightarrow \infty} {\epsilon}^2 {\epsilon}^{-\omega}\left |  \frac{{(({1-\epsilon}^4)e^{\pm i \pi})}^{1+j+\frac{i \tau j}{2}}}{{(({1-\epsilon}^4)e^{\pm i \pi})}^{1+j+1+\frac{i \tau (j+1)}{2}}}\frac{2^{(1+i\tau (j+1))}}{2^{(1+i\tau j)}}\frac{\Gamma(2 + 2(j+1))\Gamma(\frac{1}{2}){(j+1)}^{-1/2}\Gamma(m+1+j)\Gamma(1-m+j)}{\Gamma(2+2j)\Gamma(\frac{1}{2})j^{-1/2}\Gamma(m+2+j) \Gamma(2-m+j)} \right |  \\
\times \left | \left ( {\frac{(1-{\epsilon}^2) e^{\pm i \pi}}{{\epsilon}^2 +1}}\right )^{(1+\frac{i \tau (j+1)}{2} + j+1)-(1+\frac{i  \tau j}{2} + j)} \right |  \times \left | \left ( {\frac{2}{{\epsilon}^2 +1}} \right ) ^{(-\frac{1}{2}-\frac{i \tau (j+1)}{2}+m)-(-\frac{1}{2}-\frac{i \tau j}{2}+m)} \right |  \\
\times  \left | \left ( {\frac{2{\epsilon}^2}{{\epsilon}^2 + 1}} \right ) ^{(-m -\frac{i \tau (j+1)}{2}-\frac{1}{2})-(-m -\frac{i \tau j}{2}-\frac{1}{2})} \right | = \\
  \lim\limits_{j \rightarrow \infty}  \left | \frac{{\epsilon}^2 {\epsilon}^{-\omega} 2^{-\omega}}{((1-{\epsilon}^4)e^{\pm i \pi})^{(1-\frac{\omega}{2})}}   \frac{(2j+3)(2j+2)\Gamma(2j+2)\Gamma(m+1+j)\Gamma(1-m+j)}{\Gamma(2j+2)(m+1+j)(1-m+j)\Gamma(m+1+j)\Gamma(1-m+j)} \right |  \times \\ \left |  \frac{{((1-{\epsilon}^2 )e^{\pm i \pi})}^{(1-\frac{\omega}{2})}}{{({\epsilon}^2 +1)}^{(1-\frac{\omega}{2})}}\frac{2^{\frac{\omega}{2}}}{{({\epsilon}^2+1)}^{\frac{\omega}{2}}} \frac{{(2{\epsilon}^2)}^{(\frac{\omega}{2})}}{{({\epsilon}^2 + 1)}^{(\frac{\omega}{2})}} \right | = 
\frac{4{\epsilon}^2}{{({\epsilon}^2 + 1)}^2} < 1,  \; \forall |\epsilon| < 1 
\end{multline}\\[2ex]
\textbf{Case 3:} $|\epsilon = 1\; , \tau \in C, \tau = \eta + i \omega$\\
In the remaining case $\epsilon = 1$ we get $ {}_2F_1(a, b; c; 0) = 1 $ and $D^{(j, \tau j)}_{jm,  jm}(g) =1$ therefore 
$\sum\limits_{j=1}^{\infty}\frac{1}{j^k}D^{(j, \tau j)}_{jm,  jm}(g) =\sum\limits_{j=1}^{\infty} \frac{1}{j^k} $ is convergent for $k \ge 2$.\\[2ex]
Thus, we have proved by ratio test that  $\sum\limits_{j=1}^{\infty}\left |  {\epsilon}^{2(m + j + 1 + \frac{i \tau j}{2})} \right | \left|{}_2A_1(j, m, \tau, \epsilon) \right|$ is convergent. Therefore, by ($\ref{FirstSum11}$) and ($\ref{FirstSum12}$), the sum $\sum\limits_{j=1}^{\infty} \frac{D^{(j, \tau j)}_{jm,  jm}(g)}{j^k}, k \ge 2 $ is absolute convergent and also convergent. 
By the same argument as at the end of the Lemma 1 it is clear that by construction the limit of the sum is the function on $SL(2,C)$. \\[2ex]

$\square$

\section{ Square Integrability}
\label{sec:SquareIntegrability}

\textbf{Theorem 1 } 
The limit of the sum $\sum\limits_{j=1}^{\infty} \frac{D^{(j, \tau j)}_{jm, jm}(g)}{j^k}, k\ge 2, \tau \in C, j, m \in Z$ is a square integrable function for all $\tau \neq i, -i$ and fixed m.\\[2ex]
\textbf{Proof:}\\[2ex]
Both sums are convergent  for each fixed m by Lemma 2. Let $f(g)$ be the sum limit. Consider the inner product integral:
\begin{equation}
\label{SquareIntegrability2}
<f(g), \overline{f(g)}> \;\; =\int\limits_{SL(2,C)}f(g) \overline{f(g)} dg =  \int\limits_{SL(2,C)}dg \; \left (\sum\limits_{j=1}^{\infty} \frac{D^{j, \tau j}_{jm, jm}(g)}{j^k} \right ) \left (\sum\limits_{j'=1}^{\infty}\frac{\overline{ D^{j', \tau j'}_{j'm', j'm'}(g)}}{j^k} \right )
\end{equation}
The Lorentz matrix coefficients square integrability and orthogonality provides us the following equality $\cite{Clebsch-Gordan1}$, formula 9:
\begin{equation}
\label{MatrixCoeffOrtho}
\int\limits_{SL(2,C)} D^{(n_1, \rho_{1})}_{j_1q_1, {j'}_1{q'}_1}(g))\overline{ D^{(n_2, \rho_{2})}_{j_2q_2, {j'}_2{q'}_2}(g))} \; dg = {\delta}_{n_1 n_2} \frac{\delta( \rho_1 - \rho_2 )}{{n_1}^2 + {{\rho}_1}^2} \delta_{(j_1q_1), (j_2 q_2)}\delta_{({j'}_1 {q'}_1), ({j'}_2{q'}_2)}
\end{equation}
where, $n_1, n_2 \in Z, {\rho}_1, {\rho}_2 \in C$.
From  ($\ref{SquareIntegrability2}$), and ($\ref{MatrixCoeffOrtho}$) we immediately obtain:
\begin{equation}
<f(g), \overline{f(g)}> \; \; = \int\limits_{SL(2,C)}f(g) \overline{f(g)} dg =\sum\limits_{j=1}^{\infty}\frac{1}{j^k(j^2 + {\tau}^2j^2)} 
\end{equation}
Here we used the convergence of the sums for $k \ge 2$ proved in Lemma 2 to swap the integral with the sum. 
Since the sum on the right is convergent for any $k > 0$, the integral exists for all values $\tau \neq i, -i$.\\[2ex]
$\square$\\[2ex]
\textbf{Theorem 2 } $\quad$ 
The limit of the sum $\sum\limits_{j=1}^{\infty}\sum\limits_{|m| \le j}\sum\limits_{|n| \le j} \frac{D^{(j, \tau j)}_{jm, jn}(g)}{j^k} = \sum\limits_{j=1}^{\infty}\sum\limits_{|m| \le j} \frac{D^{(j, \tau j)}_{jm, jm}(g)}{j^k}, k \ge 2, \tau \in C$ is square integrable.\\[2ex]
\textbf{Proof:}\\[2ex]
The sum is convergent as proved in the Lemma1. Let $f(g)$ be the sum limit. Consider the inner product integral:
\begin{equation}
\label{SquareIntegrability1}
<f(g), \overline{f(g)}>\;\; = \int\limits_{SL(2,C)}f(g) \overline{f(g)} dg =  \int\limits_{SL(2,C)}dg \; \left (\sum\limits_{j=1}^{\infty}\sum\limits_{|m| \le j} \frac{D^{j, \tau j}_{jm, jm}(g)}{j^k} \right ) \left (\sum\limits_{j'=1}^{\infty}\sum\limits_{|m| \le j}\frac{\overline{ D^{j', \tau j'}_{j'm', j'm'}(g)}}{j^k} \right )
\end{equation}
By using matrix coefficients orthogonality ($\ref{MatrixCoeffOrtho}$), and by changing the order of sum and integral, since the sums are convergent by Lemma1, we obtain:
\begin{equation}
<f(g), \overline{f(g)}>\;\; = \int\limits_{SL(2,C)}f(g) \overline{f(g)} dg =\sum\limits_{j=1}^{\infty} \frac{2j+1}{j^k(j^2 + {\tau}^2j^2)} 
\end{equation}
The sum on the right is convergent for any $k\ge 1$, therefore the function $f(g)$ is square integrable. We had to require $k\ge 2$, since only for $k\ge 2$ the sums in Lemma 1 are convergent and we allowed us to change the order of integration and summation.\\[2ex]
$\square$\\[3ex]
\textbf{Lemma 3} $\quad$ Let  $\psi(g) \in L_2(SL(2,C)), g \in SL(2,C)$ be a square integrable function and $c_{jmm}$ being its Fourier transform coefficients:
$c_{jmm} = \int\limits_{SL(2,C)} \psi(g) \frac{\overline{D^{(j, \tau j)}_{jm,jm}(g)}}{j^k} \; dg, k \ge 2$ then the sum $\sum\limits_{j=1}^{\infty}c_{jmm}$ is convergent.\\[2ex]
\textbf{Proof:}\\
\begin{equation}
\label{SquareIntegrability1}
\sum\limits_{j=0}^{\infty} c_{jmm} = \sum\limits_{j=0}^{\infty} \; \int\limits_{SL(2,C)} \psi(g)\frac{ \overline{D^{(j, \tau j)}_{jm, jm}(g)}}{j^k} \; dg = \int\limits_{SL(2,C)} \psi(g)  \sum\limits_{j=0}^{\infty} \frac{\overline{D^{(j, \tau j)}_{jm,jm}(g)}}{j^k}\; dg
\end{equation}
However by Theorem 1 the sum $\sum\limits_{j=1}^{\infty}\frac{{D^{(j, \tau j)}_{jm,jm}(g)}}{j^k}$ is square integrable for $k\ge 2$, and therefore the $\sum\limits_{j=0}^{\infty}\frac{\overline{D^{(j, \tau j)}_{jm,jm}(g)}}{j^k}$ is also square integrable and converges to the square integrable function $\phi(g) \in L_2(SL(2,C))$. Therefore the integral $\int\limits_{SL(2,C)}\psi(g) \phi(g) \; dg$ in ($\ref{SquareIntegrability1}$) is convergent and so the sum $\sum\limits_{j=1}^{\infty}c_{jmm}$.\\[2ex]
$\square$\\[2ex]

\section{ The Y-Map: $L_2(SU(2)) \rightarrow F(SL(2,C))$}
\label{sec:YMap}

We are going to prove that any square integrable function  $\phi(u)$ on $SU(2)$ can be mapped to a function $\psi(g)$ on $SL(2,C)$ (not necessary square integrable), by using the solution of the $SU(2)$ simplicity constraints $\cite{EPRL}$: ($k = j, \rho = \tau j, j \in Z, \gamma \in C$) in  the following manner:
\begin{equation}
\label{SU(2)YMap}
\phi(u) \rightarrow \psi(g) = \sum \limits_{j=|p|}^{\infty}\sum\limits_{|m| \le j}c_{|p|m}^{\frac{j}{2}} D_{jm, jm}^{(j, \tau j)}
\end{equation}
, where $c_{|p|m}^{\frac{j}{2}}$ is $\phi(u)$'s Fourier transform:
\begin{equation}
\label{SU(2)-Fourier}
c_{|p|m}^{\frac{j}{2}} = (j+1)^{\frac{1}{2}}\int \limits_{SU(2)} \phi(u) \overline{D_{|p|m}^{\frac{j}{2}}(u)} \; du
\end{equation}
, where $p, j, m, n \in Z, \; \gamma \in C, \; u \in SU(2), \; g \in SL(2,C)$
Note that the parameter $j$ in $SL(2,C)$ matrix coefficients is an integer, while the parameter of the $SU(2)$ in $c_{|p|m}^{\frac{j}{2}}$ is half-integer.
\\[4ex]
\textbf{Theorem 3 -  SU(2)-Y-Map Existence Theorem: }\\[2ex]  
The sum $ \phi(u) \rightarrow \psi(g) = \quad \sum\limits_{j=|p|}^{\infty}\sum\limits_{|m| \le j} c^{\frac{j}{2}}_{|p|m} D^{(j, \tau j)}_{jm, jm}(g) $ is convergent. \\[2ex]
\textbf{Proof:}\\[2ex]
By Paley-Wiener Theorem ($\cite{Ruhl}$ page 60, 91, see also $\cite{Gelfand}$) the Fourier transform $ c^{\frac{j}{2}}_{|p|m}$ satisfies the following asymptotic inequality:
$ \forall k \in N, k \ge 1$ or we can rewrite it as:
\begin{equation}
\label{asymptotic3}
|c^{\frac{j}{2}}_{|p|m}| \le \frac{C_k}{|j|^k}
\end{equation}
which means that the Fourier transform is a fast dropping function and decreases faster than any polynomial of power $k$, where $C_k$ is a constant depending on k only. Therefore
\begin{multline}
\sum\limits_{j=|p|}^{\infty}\sum\limits_{|m| \le j} c^{\frac{j}{2}}_{|p|m} D^{(j, \tau j)}_{jm, jm}(g) \le  \sum\limits_{j=|p|}^{\infty}  \sum\limits_{|m| \le j} |c^{\frac{j}{2}}_{|p|m} D^{(j, \tau j)}_{jm,  jm}(g)| \le \sum\limits_{j=|p|}^{\infty}  \sum\limits_{|m| \le j} |\frac{C_k}{|j|^k} D^{(j, \tau j)}_{jm,  jm}(g)| \\
\le |C_k|\sum\limits_{j=|p|}^{\infty}  \sum\limits_{|m| \le j} |\frac{ D^{(j, \tau j)}_{jm,  jm}(g)}{j^k}|
\end{multline}
and the last sum is convergent by Lemma1: $\sum\limits_{j=1}^{\infty}  \sum\limits_{|m| \le j} \frac{ D^{(j, \tau j)}_{jm,  jm}(g)}{j^k}$ is absolute convergent and therefore also convergent for any $k \in N, k \ge 2$. In Appendix we also provide the graph from the numerical calculations by using MPMath Python program to demonstrate visually convergence proved in Lemma 1. \\
The limit is a function on $SL(2,C)$  since each $g \in SL(2,C)$ we map to the sum limit and the limit is unique by construction.\\[4ex]
$\square$\\[2ex]
SU(2)-Y-Map Existence Theorem establishes a map from the space of square integrable functions on $SU(2)$ to the space of  functions (not necessarily square integrable) on $SL(2,C)$.\\[2ex]

Acknowledgment:
I would like to specially thank Michael Bukatin for multiple fruitful and challenging discussions.

\section{ Discussion }
\label{sec:Discussion}
We proved the convergence the sums of the $SL(2,C)$ principal series matrix coefficients when summed not over all parameters but only over the solutions of the simplicity constraints. We also proved that the limit is a square integrable function on $SL(2,C)$. We used those facts in the proof of the main result of this paper - the Y-Map convergence provided in the Theorem 3.  The Y-Map is the map from the space of the square integrable functions on $SU(2)$ to the space of functions on $SL(2,C)$. \\[2ex]

\section{Appendix B Numerical Demonstration}
\label{sec: NumericalConfirmation}
The numeric calculation with MPMath Python program $\cite{mpmath}$ provides the following results for the $SL(2,C)$ matrix coefficients sum convergence:
\begin{equation}
\label{SL2CMatrix3}
\sum\limits_{j=1}^{\infty}  \sum\limits_{|m| \le j} \frac{D^{(j, \tau j)}_{jm, jm}(g)}{j^k} = \sum\limits_{j=1}^{\infty}  \sum\limits_{|m| \le j}\frac{1}{j^k}{\epsilon}^{2(m + j + 1 + \frac{i \tau j}{2})} {}_2F_1( j + 1 + \frac{i \tau j}{2}, m+j+1; 2j+2; 1-{\epsilon}^4)
\end{equation}
with $\epsilon = 0.5$, Immirzi $\tau = 0.127$, for j limits from 0 to 200. The python code and the graph are as follows:\\[2ex]
For k = 2\\[2ex]
\includegraphics[width = 150mm, height=40mm]{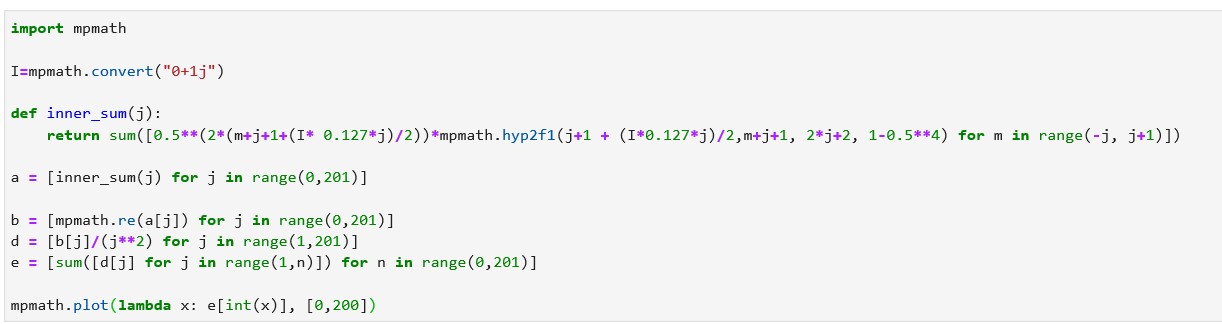}\\
\includegraphics[width = 60mm, height=40mm]{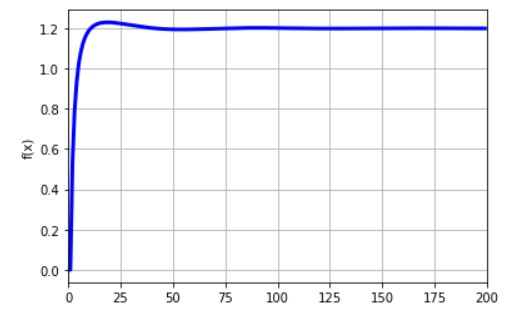}\\
For k = 6\\[2ex]
\includegraphics[width = 150mm, height=40mm]{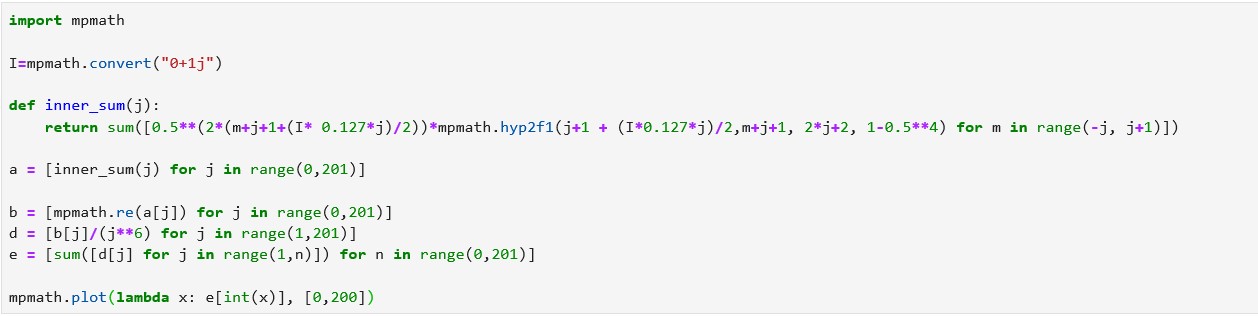}\\
\includegraphics[width = 60mm, height=40mm]{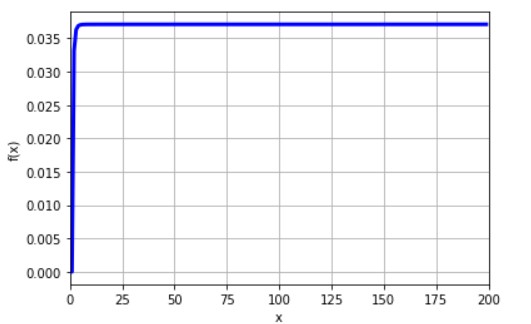}\\

\end{document}